\documentclass{elsart}
\usepackage{natbib}
\usepackage{graphics}
\usepackage{epsfig}

\begin{document}
\runauthor{Tiesinga}
\begin{frontmatter}

  \title{Attentional modulation in layer 4 of the visual cortex could
  be mediated by interneurons with complex receptive field
  characteristics}

  \author{Paul H. E. Tiesinga and Calin I. Buia}

  \address{Computational Neurophysics Laboratory, Department of
    Physics \& Astronomy, University of North Carolina at Chapel Hill,
    244 Phillips Hall, Chapel Hill NC 27599, USA.}

  \thanks[Someone]{This work was supported by 
   the University of North Carolina at Chapel Hill.}

  \thanks[CorAut]{Corresponding author, {\tt tiesinga@physics.unc.edu}}

\begin{abstract}
Many neurons in the visual cortex are orientation-selective, increase
their firing rate with contrast and are modulated by attention. What
is the cortical circuit that underlies these computations? We examine
how synchrony can be modulated by the excitability
of interneurons, in a model layer 4 network displaying contrast-invariant
orientation-tuning. We did not find parameter settings for which the
standard ring model (\cite{Somers95}), which contains only cells
with simple receptive fields (RF), behaved appropriately. Only when
interneurons with complex receptive fields were included, similar to
those found recently in cat primary visual cortex (\cite{Hirsch03}),
did the network behave appropriately. A critical feature in the 
model was that complex interneurons projected to simple interneurons
but the simple interneurons did not project back to them. The
network was switched from the non-attended state to the attended state
by increasing the depolarization of the complex interneurons. In
addition to contrast-invariant orientation tuning, the model
reproduced the following experimental results: (1) the gamma-frequency
range coherence between the estimated local field potential (eLFP) and
spike trains of excitatory cells was higher in the attended state than
in the non-attended state, but the firing rates of the excitatory
cells did not vary between states; (2)
the gamma-frequency-range power in the eLFP increased with
contrast. The model predicts that there are two populations of
inhibitory cells, one with complex RF characteristics whose firing
rate increases with attention and the other with simple RF
characteristics whose firing rate decreases with attention.
\end{abstract}

\begin{keyword}
synchronization, noise, gamma oscillations
\end{keyword}

\end{frontmatter}

\section{Introduction}

Inhibitory interneurons make up 15\% to 20\% of all cortical neurons
and have been classified into distinct groups based on their
morphology, the complement of calcium-binding proteins they express
and their physiological properties \citep{Markram04}. It is
commonly thought that interneurons function primarily as a brake on
recurrent excitation, thereby preventing epilepsy, but recent work
suggests that they may also play a direct role in cortical information
processing \citep{Santha04,Eche05,Llinas05}. For instance, in the ring model
\citep{Ben95,Somers95},
orientation-selectivity emerges in part because inhibitory neurons
sharpen the weakly orientation-tuned inputs that cortical neurons receive
from the lateral geniculate nucleus (LGN), whereas recent models
suggest that attention may be mediated by the synchrony of inhibitory
neurons \citep{Fries01,Bichot05,Tiesinga05,Buia06}. It is not clear
whether both attention 
and sharpening of tuning are achieved by the same class of
interneurons because it has not yet been possible to conclusively
link the classification of interneurons to their function in the
cortical circuit. We use computational models to investigate what
types of interneurons are necessary for attentional modulation of
orientation-selective neurons in layer 4 of the visual cortex. Our
hypothesis is that spatial attention is mediated by a depolarization
of interneurons, which in turn synchronize the cortical network
\citep{Buia06}. In our simulations synchrony could not
be modulated effectively in a ring model containing only simple
inhibitory cells, whereas when the network included complex inhibitory
cells projecting to simple inhibitory cells, synchrony could be
modulated effectively. Evidence for complex and simple inhibitory
cells was recently obtained in the visual cortex of the anesthetized
cat using intracellular recordings \citep{Hirsch03}. In addition,
anatomical studies have revealed evidence for various kinds of
complex-like inhibitory neurons projecting preferentially to other
inhibitory neurons \citep{Gonchar99,Gonchar03} and receiving
top-down inputs from other cortical areas or from neurons in the basal
forebrain \citep{Freund92}. 
 
The simulations presented here provide predictions for how the firing
rates of two types of interneurons are differentially modulated with
attention. This information is not only useful for the {\it in vivo}
identification of interneuron type using multi-electrode recordings,
it may also help to correlate their functional role with their
anatomical and biochemical characteristics using intracellular recordings. 

\section{Methods}
\subsection{Cortical  models}
\begin{figure}
\centering
\epsfig{file=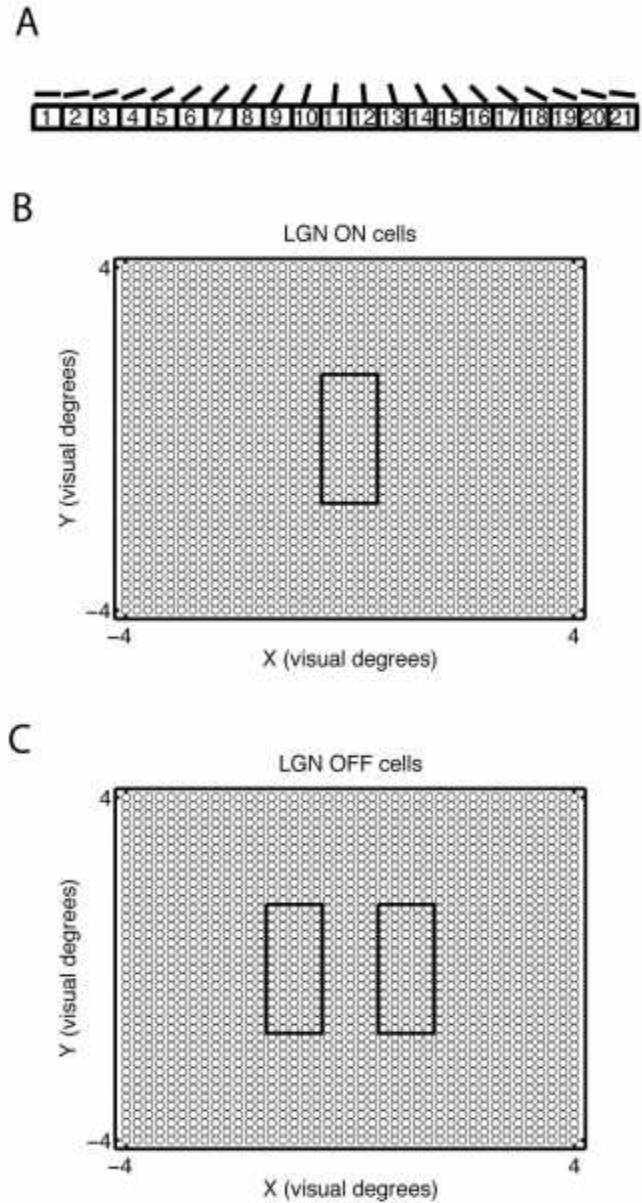, width=3.5in}
\caption{Generation of orientation selectivity in the LGN to cortex
  projection. (A) There were 21 columns with excitatory and inhibitory
  neurons, whose preferred orientation smoothly varied from horizontal
  ($0^{\circ}$) via vertical ($90^{\circ}$) to almost a $180^{\circ}$. The
  preferred orientation of the rightmost column is close to that of
  the leftmost columns, hence these can be identified as neighbors,
  giving the model a ring architecture. (B,C) There were 1681 LGN-ON
  and 1681 LGN-OFF cells laid out on a 41 by 41 grid spanning 8 by 8
  visual degrees. An excitatory cortical cell with a receptive field
  center at $X=0$ and $Y=0$ and a preferred orientation of $90^{\circ}$,
  would receive inputs from 12 ON cells located in the rectangle in
  (B) and inputs from 12 OFF cells located in either of the rectangles
  shown in (C). These rectangles (subfields) were rotated and shifted
  appropriately for a neuron with a different receptive field center
  and preferred orientation.
}
\label{fig1}
\end{figure}

The ring model consisted of $N_c=21$ columns, each comprised of 21
inhibitory and 84 excitatory cells (total: 1764 excitatory cells, 441
inhibitory cells). These numbers are representative for a hypercolumn
in cat visual cortex \citep{Somers95}. The preferred orientation
varied smoothly from column to column: for the $i$th column it was $180
^{\circ} (i-1)/N_c$ (Figure \ref{fig1}A). Since the left-most column and the
right-most column have similar orientation preferences, the model is
referred to as a ring model. The LGN was represented using 1681 ON and
1681 OFF cells laid out on two overlapping 41 by 41 grids spanning 8
by 8 degrees of the visual field with a center to center spacing of
0.2 degrees in both the $x$ as well as the $y$ direction (Figure
\ref{fig1}B and C). The grids of ON and OFF cells were fully
overlapping. The LGN 
inputs to each cortical neuron came from 3 rectangular subfields, each
1 by 3 degrees in size, with their long axis oriented along the
preferred orientation of the neuron (Figure \ref{fig1}B and C). The
OFF, ON and OFF subfields were arranged from left to right with their
long axes parallel and with a center-to-center distance of 1 degree. For
each cortical neuron, an appropriate number (see Table \ref{tab1}) of LGN-OFF
cells were randomly selected out of the area covered by the OFF
subfield (Figure \ref{fig1}C) and were connected to the neuron. A similar
procedure was applied to the LGN-ON cells, which were selected out of
the ON subfield (Figure \ref{fig1}B).  

\begin{table}
\centering
\begin{tabular}{|l|c|c|c|c|c|c|c|}
\hline
Type & $g$ & $N_{syn}$ & $\tau_{syn}$ & $\delta$ & $\sigma_{\delta}$
& $\sigma$  & $mr$ \\
  & ($\mu\textrm{S/cm}^2$) &  & (ms) & (ms) & (ms) & (rad) & (rad) \\
\hline
LGN$\rightarrow$E & 14 & 12 & 5 & 10 & 5 & -- & -- \\
\hline
LGN$\rightarrow$SI & 8 & 8 & 5 & 5 & 3 & -- & -- \\
\hline
E$\rightarrow$E & 6 & 36 & 5 & 3 & 1 & 0.1 & 0.7 \\
\hline
E$\rightarrow$SI & 3 & 56 & 5 & 3 & 1 & 0.1 & 0.7 \\
\hline
SI$\rightarrow$E & 12 & 24 & 10 & 3 & 1 & 1 & 1.5 \\
\hline
SI$\rightarrow$SI & 5 & 8 & 10 & 3 & 1 & 1 & 1.5 \\
\hline
\end{tabular}

\caption{The parameter settings for synaptic connections in the
  standard ring model. $g$ is the unitary strength of the synapse,
  $N_{syn}$ is the number of synapses, $\tau 
  _{syn}$ is the synapstic time constant, $\delta$ is the mean
  conduction delay, $\sigma _{\delta}$ is the standard deviation of
  the conduction delay}
\label{tab1}
\end{table}

The relative probability $P$ for a connection between a presynaptic
neuron $i$ (preferred orientation: $\theta_i$) and a postsynaptic neuron $j$
(preferred \mbox{orientation: $\theta_j$}) depended on the difference
between their 
preferred orientations, $P=f(\theta_i-\theta_j,\sigma,mr)$. Here
$f(\theta_i-\theta_j,\sigma,mr)=\exp(-x^2/2\sigma^2)$ for $|x|<mr$ and
$f=0$ for $|x|>mr$, $\sigma$ is the 
orientation tuning width and $mr$ stands for maximum range and is the
orientation difference beyond which no connections are made. For each
cortical neuron, an appropriate number of presynaptic neurons were
chosen with relative probability $P$ and connected to the neuron. The
values used for the model are summarized in Table \ref{tab1}.  

\begin{figure}
\centering
\epsfig{file=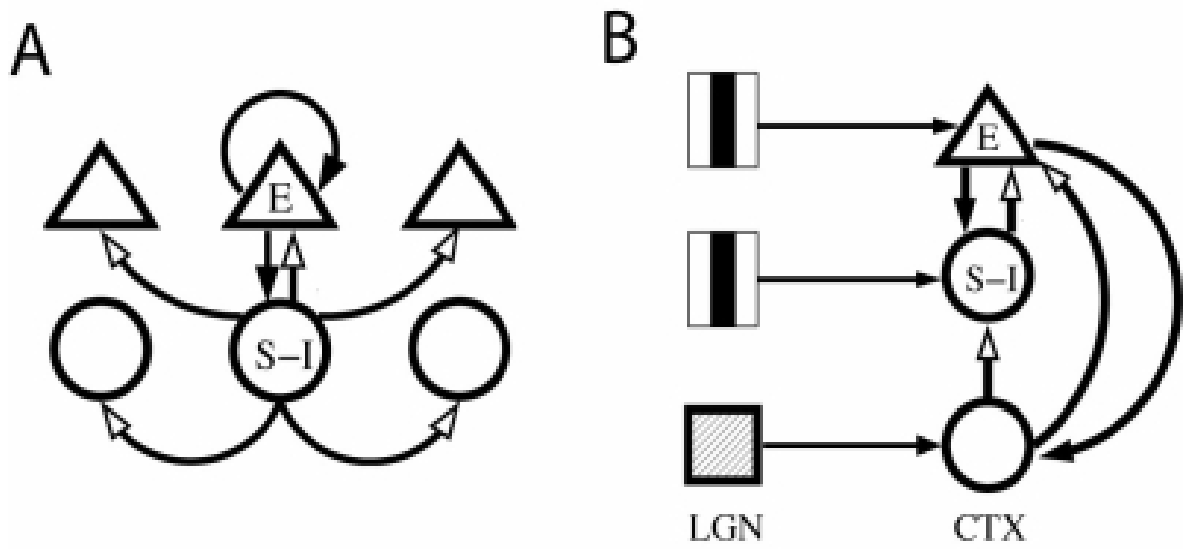, width=5in}
\caption{The local circuit. (A) The standard ring model consisted of
  simple excitatory (E) and inhibitory (SI) cells, each with a
  receptive field comprised of parallel subfields arranged in an
  OFF-ON-OFF configuration. The excitatory cells only projected to
  other excitatory and inhibitory cells with a similar orientation
  preference (filled arrows), whereas the inhibitory neurons projected
  to other neurons with orientation preferences in a broader range
  around that of the home-column (open arrows). (B) In the modified
  network an additional inhibitory cell type (CI) with complex
  receptive field characteristics was introduced. Only the numerically
  dominant projections are shown in the graph. These neurons received
  inputs from LGN-ON and LGN-OFF cells with overlapping receptive
  fields centers. The CI cells projected to the simple excitatory and
  inhibitory cells, regardless of their preferred orientation. A CI
  cell received excitatory inputs with equal probability from all
  preferred orientations. Only a few SI cells provided inputs to the
  CI cells.
}
\label{fig2}
\end{figure}

For the 'complex' version of the model, referred to as the complex
ring model, 11 inhibitory cells with complex RFs were added to each
column. These cells were labeled by the preferred orientation of the
column to which they were assigned. They received LGN inputs from ON and OFF
cells in a 3 by 3 degrees area centered on the neuron's receptive field
center (Figure \ref{fig2} B). The intracortical connections between complex and
simple cells did not depend on the difference between the preferred
orientation of the pre- and postsynaptic neuron (formally: $\sigma=\infty$ and
$mr=\pi/2$). There was an asymmetry in the connection between
inhibitory neurons 
with a simple receptive field (SI) and those with a complex receptive
field (CI). Most inhibitory inputs to CI cells came from other CI
cells, but a significant fraction of inhibitory inputs to SI cells
came from CI cells (Table \ref{fig2}). The excitatory cells received inputs
from both CI and SI cells.

\begin{table}
\centering
\begin{tabular}{|l|c|c|c|c|}
\hline
Type & $g$ & $N_{tsyn}$ & $\sigma$ & $mr$ \\
  &  &  & (rad) & (rad) \\
\hline
LGN$\rightarrow$E & 14 & 12 & -- & -- \\
\hline
LGN$\rightarrow$SI & 8 & 8 & -- & -- \\
\hline
LGN$\rightarrow$CI & 8 & 4 ON, 2 OFF & -- & -- \\
\hline
E$\rightarrow$E & 6 & 36 & 0.1 & 0.7 \\
\hline
E$\rightarrow$SI & 1 & 56 & 0.1 & 0.7 \\
\hline
E$\rightarrow$CI & 1 & 56 & $\infty$ & $\pi/2$ \\
\hline
SI$\rightarrow$E & 20 & 19.6 & 1 & 1.5 \\
\hline
CI$\rightarrow$E & 20 & 4.4 & $\infty$ & $\pi/2$ \\
\hline
SI$\rightarrow$SI & 4 & 26.25 & 1 & 1.5 \\
\hline
SI$\rightarrow$CI & 4 & 0.76 & $\infty$ & $\pi/2$ \\
\hline
CI$\rightarrow$SI & 4 & 13.75 & $\infty$ & $\pi/2$ \\
\hline
CI$\rightarrow$CI & 4 & 39.24 & $\infty$ & $\pi/2$ \\
\hline
\end{tabular}
\caption{The parameter settings for the complex ring model. $g$ is the
  unitary strength and $N_{tsyn}$ is the number of synapses on the target. The
  conduction delays and the synaptic time constants are as in Table
  \ref{tab1}.} 
\label{tab2}
\end{table}

\subsection{Calculation of LGN spike trains used as input to the
  cortical neurons.}

The LGN cells were represented by non-separable spatiotemporal
filters, which were defined in terms of a Gaussian spatial filter
$F(x,y|\sigma,K)=(K/2\pi\sigma^2) \times \exp(-(x^2+y^2)/2\sigma^2)$ and 
an alpha-type temporal filter $G(t|\tau)=(1/\tau)\exp(-t/\tau)$. Specifically,
$F_{st}=F(x,y|\sigma_c,K_c)G(t,\tau_c)-F(x,y|\sigma_s,K_s)G(t-\delta,\tau_s)$.
Here $x$ and $y$ are 
defined with respect to the receptive field center of the LGN
cell. Parameters values are $K_c/K_s=17/16$, $\sigma_c=0.17^{\circ}$,
$\sigma_s=0.53^{\circ}$, $\tau_c=10$ms, $\tau_s=20$ms and the delay
between the surround and center response is $\delta=3$ms. The firing
rate of the ON cells is $R_{ON}(t)=[R_0+c_{lin}F_{sc}(F_{st}*s)]_+$
and for the OFF cells it is $R_{OFF}(t)=[R_0-c_{lin}F_{sc}(F_{st}*s)]_+$. In
these expressions $R_0=15$Hz is the baseline firing rate of 
thalamic neurons \citep{Somers95}, $c_{lin}$ is the {\it linear} contrast, a
scaling factor between 0 and 1 (0 and 100\%), $s$ is the stimulus
waveform (see below), $F_{st}*s$ is the scalar filter output
determined as a sum across space and a convolution in time, $F_{sc}$
is a scaling factor such that $F_{sc}(F_{st}*s)$ is 50 Hz during the
sustained part of the response to the 
vertical bar stimulus (see below), and $[x]_+$ denotes rectification. The
spike trains for LGN neurons are obtained as a Poisson process with a
time-varying rate given by $R_{ON}(t)$ or $R_{OFF}(t)$ for ON and OFF cells,
respectively. The filter is 64 ms long (sampling rate is 1 kHz),
shifting the stimulus onset in the filtered temporal waveform by 32
ms. Hence, in combination with the 5 to 10 ms axonal conduction delay
(see Table \ref{tab1}), stimulus-onset reached cortical neurons after a 40 ms
delay.

\subsection{Stimulus generation}

The stimulus was represented as a spatio-temporal waveform $s(x,y,t)$,
with a pixel value of zero representing the gray background. A
vertical bar of 1 by 3 degrees was generated by setting the values of
$s(x,y,t)$ equal to one for $t_1<t<t_2$, $-0.5^{\circ}<x<0.5^{\circ}$ and
$-1.5^{\circ}<y<1.5^{\circ}$ during either a 300 ms long period starting at
$t_1=250$ms or 400ms, or a 600ms period starting at $t_1=800$ms. We
used a temporal resolution of 1 ms and a spatial resolution of
$0.1^{\circ}$ for the stimulus matrix.  

\subsection{Neuron and synapse models}

The neurons were represented by Hodgkin-Huxley-style models. For the
inhibitory neuron, we used the model in \citep{Wang96} and
for the excitatory neuron, we used the model in
\citep{Golomb97,Golomb98}. The AMPA-type excitatory and fast GABA-type 
inhibitory synapses were also taken from \citep{Golomb97,Golomb98}. The
implementation details and equations for the cell 
and synapse model were given in a previous publication and are not
repeated here \citep{Buia05,Buia06}. The
only single-neuron parameters varied during the course of the
simulations reported here are the level of depolarizing current $I_{CI}$,
$I_{SI}$ and $I_E$ to the complex inhibitory, simple inhibitory and excitatory
neurons, respectively, and for the excitatory cells, the maximum
conductance $g_{Kslow}$ (standard value 0.075$\textrm{mS/cm}^2$) of
slow potassium current responsible for 
adaptation \citep{Golomb97,Golomb98}, and the decay time
of the corresponding kinetic variable (standard value:
$\tau_z=75$ms). NMDA synapses were implemented according 
to \citep{Golomb97,Golomb98} with a time-scale of 149 ms. We fixed 
their unitary strength to 32\% of the strength of the AMPA
conductance. 

\subsection{Calculated quantities.}

Spike times were determined as the times the voltage crossed 0 mV from
below. The firing rate was the number of spikes produced during the
stimulus period divided by its duration. The rates were averaged
across all neurons of a given type (excitatory, simple inhibitory,
complex inhibitory) in a column. A tuning curve was constructed by
plotting this firing rate as a function of the preferred orientation
of the column. We performed the same analysis on the synaptic inputs
that a cortical neuron receives from LGN neurons and other cortical neurons. 

For spectral analysis we used the multi taper routines implemented in
the Chronux MATLAB toolbox \citep{Mitra99,Jarvis01}. First, a spike
time histogram was constructed for each neuron 
type by calculating the number of spikes in a 1 ms wide
bin. Histograms were normalized by the number of neurons and the bin
width in seconds, yielding the time-varying average firing rate
expressed in Hz. The power spectrum density of the histograms was
calculated using the Chronux routine $mtspectrumpb$ with a spectral
bandwidth $NW=3$, averaged over 5 tapers. In this routine, the length of
the time series was increased to twice the next integer power of two
by zero-padding. Two types of coherencies were calculated using the
routine $coherencypb$ with the same parameter values as used for the
power spectrum. The coherence between histograms of two different
types of neurons was calculated. In addition, the coherence between
the histogram of the complex inhibitory cells and the spike train of
an excitatory neuron was determined and averaged across 100 spike
trains randomly picked among the 1764 excitatory cells (we excluded
spike trains with no spikes, since these led to division by zero
errors).  

\begin{figure}
\centering
\epsfig{file=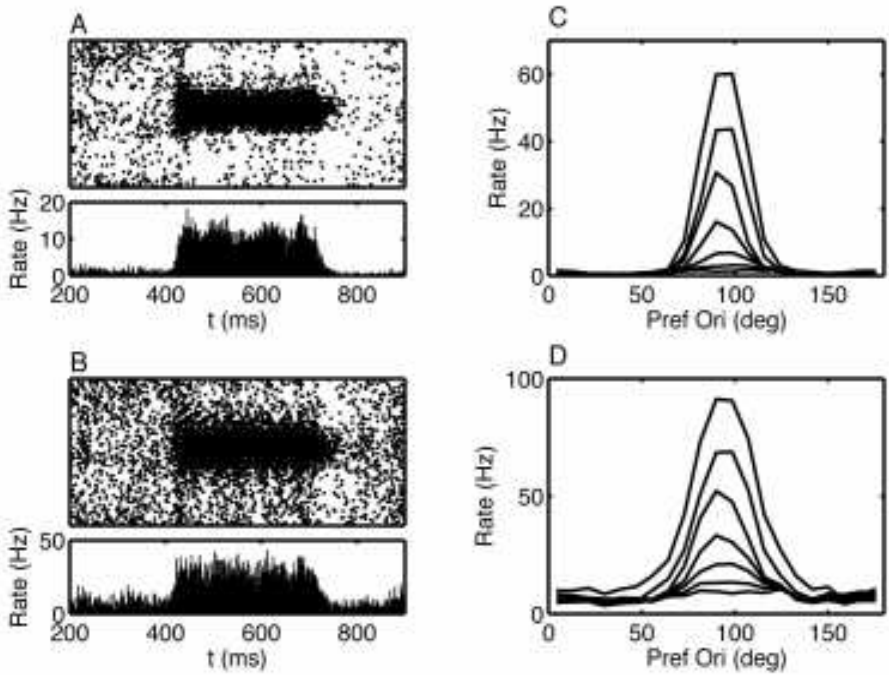, width=5in}
\caption{Orientation-selective response of the standard ring model to
  a vertical bar. A 1 by 3 degrees vertical bar was presented at full
  contrast between 400 ms and 700 ms.  (A, B) In each panel, (top) the
  rastergram and (bottom) the histogram are shown for all the (A)
  excitatory and (B) inhibitory neurons. (C,D) The mean firing rate
  during the stimulus period as a function of the preferred
  orientation of the column, averaged across all (C) excitatory and
  (D) inhibitory neurons in the column. From bottom to top, the tuning
  curves are for 0, 5, 10, 15, 25, 50, and 100 \% linear contrast.
  The driving currents (expressed in $\mu\textrm{A/cm}^2$) were
  $I_E=-0.6$, $I_{SI}=-0.1$. 
}
\label{fig3}
\end{figure}

\section{Results}

The standard ring model produced an orientation selective
response. The rastergram and histogram for the excitatory and
inhibitory cells are shown in Figure \ref{fig3}A and B, respectively. The mean
firing rate of a column in response to a vertical bar is shown as a
function, $f_{population}(\theta_p)$ , of its preferred orientation
$\theta_p$ (Figure \ref{fig3}C, D). Experimentally, an orientation
tuning curve, $f_{neuron}(\theta_s)$, is obtained by presenting 
stimuli of different orientation $\theta_s$ to a single neuron with a specific
preferred orientation. Because of the ring symmetry and the fact that
here the firing rate of a neuron only depends on the absolute value of
the difference between its preferred orientation and the stimulus
orientation, these tuning functions are directly related to each other
according to the mathematical identity
$f_{neuron}(\theta_s|\theta_p)=f(|\theta_s-\theta_p|)=
f_{population}(\theta_p|\theta_s)$. The tuning function for 
excitatory neurons was sharper (half width at half height  (HWHH) was
$15^{\circ}$) than that for inhibitory neurons (HWHH=$20^{\circ}$). The tuning
functions were approximately contrast-invariant. Contrast in the
simulations is a linear scaling factor between zero and one, with zero
yielding a baseline rate of 15 Hz for all the LGN cells, and one
corresponding to the LGN-ON cells firing at 65 Hz in response to a 1
by 3 degrees bar covering their entire RF. This does not take into account
the nonlinearity of the contrast response function of LGN neurons
\citep{Cheng95}. These nonlinearities can be accounted for by an
appropriate scaling of the $x$-coordinate in our graphs. However, plotting
linear contrast makes it easier to distinguish the nonlinearity due to
cortical dynamics from that due to the LGN input. Furthermore, the
experimental LGN curves were obtained in response to gratings rather
than the bars used here and thus may only be qualitatively correct. 

\begin{figure}
\centering
\epsfig{file=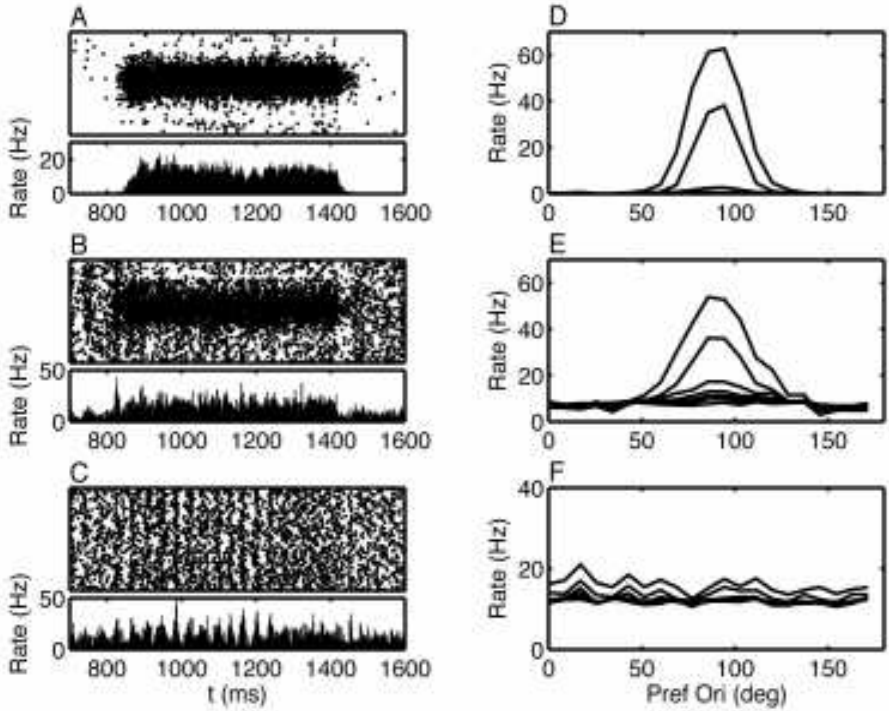, width=5in}
\caption{The response of the complex ring model in the non-attended
  condition. A  vertical bar was presented at full contrast between
  800 and 1400 ms. (A-C) We show (top) the rastergram and (bottom) the
  histogram averaged across (A) excitatory cells, (B) simple
  inhibitory cells and (C) complex inhibitory cells. (D-F) The firing
  rate in response to a vertical bar as a function of the preferred
  orientation of the column for (D) excitatory cells, (E) simple
  inhibitory cells and (F) complex inhibitory cells. From bottom to
  top, the tuning curves are for 0, 5, 10, 15, 25, 50, and 100 \%
  linear contrast. The driving currents (expressed in
  $\mu\textrm{A/cm}^2$) were $I_E=-0.6$, $I_{SI}=0.1$, $I_{CI}=0.4$.
}
\label{fig4}
\end{figure}

We added the complex inhibitory cells to the network without altering
the strength of the recurrent excitatory connections. However, in
order to obtain contrast-invariant orientation tuning it was necessary
to change the other connections (compare Table \ref{tab1} with
\ref{tab2}). Briefly, the 
number of inhibitory synapses received by each inhibitory neuron was
increased from 8 to 40, the total strength of recurrent inhibition was
increased by a factor four, and the unitary strength of the excitatory
synapse to inhibitory neurons was halved. The resulting network
activity for the non-attended condition is shown in Figure \ref{fig4}. There
were three obvious changes compared to Figure \ref{fig3}. First, the tuning was
less sharp (Figure 4D, HWHH=$17^{\circ}$ for excitatory cells and
$20.5^{\circ}$ for inhibitory cells). Second, the contrast sensitivity
had decreased 
(Figure \ref{fig4}D), as responses above 10 Hz were only obtained for contrast
values of 30\% and higher, compared with 12\% for the standard ring
model. Third, the maximum rate of inhibitory neurons had decreased
from 90 Hz to 50 Hz (Figure \ref{fig4}E). The complex cells were not
orientation-selective and their firing rate varied only weakly with
contrast (Figure \ref{fig4}F). For a 100\% contrast stimulus (Figure
\ref{fig4}C), the complex cells were weakly synchronized with an
oscillation frequency 
of about 28 Hz. For lower contrast, the complex cells were not
synchronized (see below and Figure \ref{fig8}). When the CI network is in the
non-attended condition, an increase in the level of depolarizing
current or in the amount of excitatory inputs will synchronize it (see
\citep{Tiesinga04}).

\begin{figure}
\centering
\epsfig{file=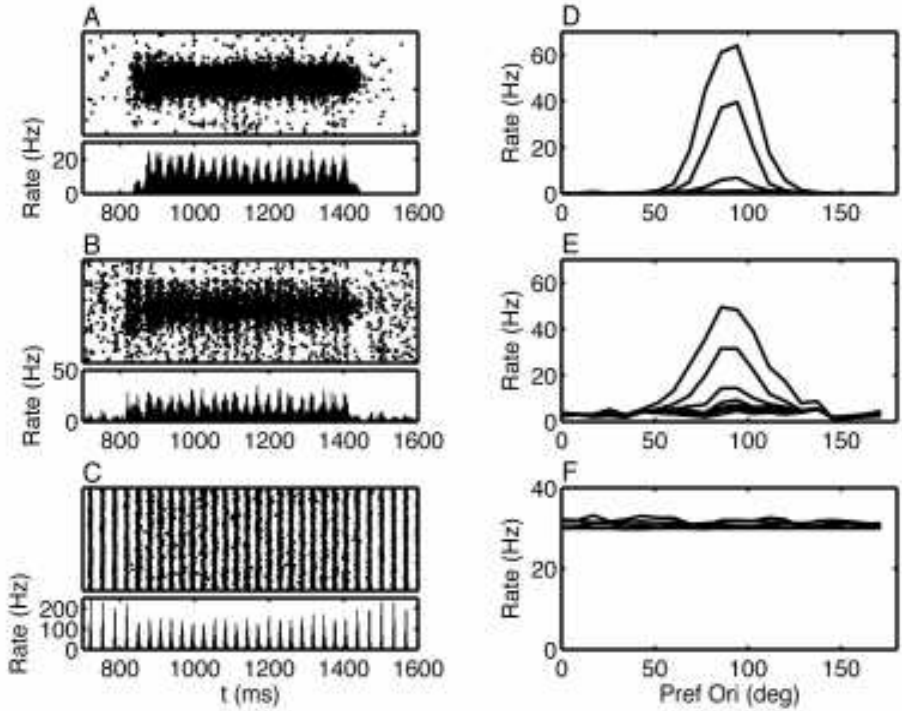, width=5in}
\caption{The response of the complex ring model in the attended
  condition.  A  vertical bar was presented at full contrast between
  800 and 1400 ms. The panels are as in Figure \ref{fig4}. The driving currents
  (expressed in $\mu\textrm{A/cm}^2$) were $I_E=-0.6$, $I_{SI}=0.1$,
  $I_{CI}=0.9$. 
}
\label{fig5}
\end{figure}

To mimic the effects of attention the depolarizing current to the CI
cells was increased from $I_{CI}=0.4$ to 0.9 $\mu\textrm{A/cm}^2$
(Figure \ref{fig5}). The CI neurons 
were synchronized in the prestimulus period and remained synchronized
during the stimulus presentation (Figure \ref{fig5}C). In response to the
oriented stimulus, the excitatory and simple inhibitory cells
increased their rate and became synchronized. The orientation tuning
functions (Figure \ref{fig5}D) were virtually identical to those in the
non-attended condition (Figure \ref{fig4}D). 

\begin{figure}
\centering
\epsfig{file=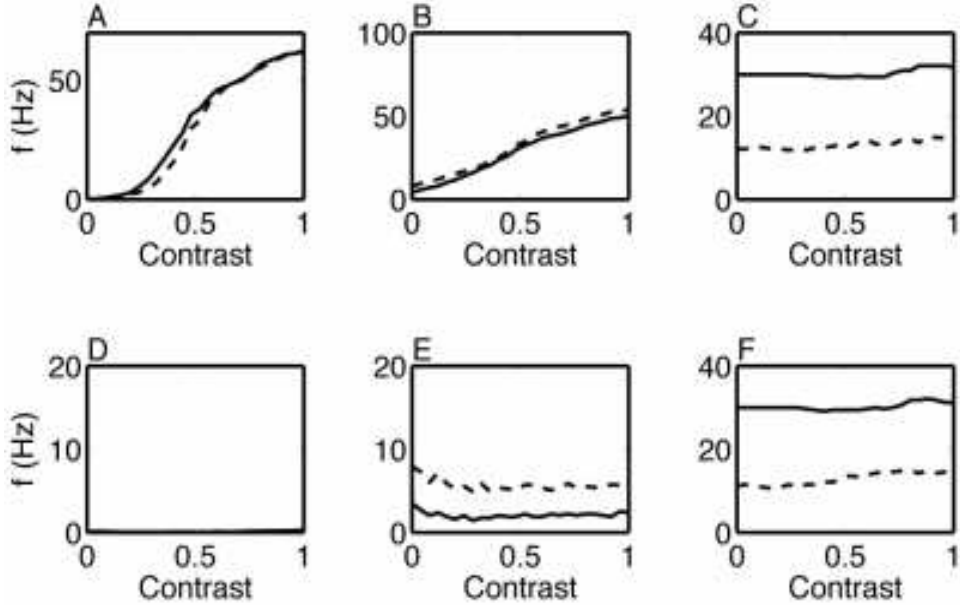, width=5in}
\caption{Attentional modulation of the contrast response functions. We
  show the CRF of neurons in the column (A-C) whose preferred
  orientation matched the stimulus orientation and (D-F) whose
  preferred orientation was orthogonal to the stimulus
  orientation. The CRFs were averaged across (A,D) all excitatory
  neurons, (B,E) simple inhibitory neurons and (C,F) complex
  inhibitory neurons. The responses are shown for a network in the
  attended (solid lines) and in the non-attended condition (dashed
  lines). 
}
\label{fig6}
\end{figure}

The effect of attention on the contrast response functions was
determined for neurons whose preferred orientation matched the
stimulus orientation (Figure \ref{fig6}A-C). For contrasts up to 48\% (at a
firing rate of 36.2 Hz), the firing rate of the excitatory cells
increased nonlinearly with contrast (Figure \ref{fig6}A). For higher contrast,
the rate of increase of the firing rate with contrast leveled off,
yielding approximately 64 Hz at 100\% contrast. Note that for even
higher LGN firing rates, corresponding to contrast values larger than
100\%, the firing rate still increased with contrast.  The CRF in the
attended condition is almost the same: the difference can be described
as a small leftward shift of 5\% with respect to the curve in the
non-attended condition. The CRF of the SI cells was reduced in the
attended condition compared with the non-attended condition (Figure
\ref{fig6}B). The firing rate of the CI cells was only weakly contrast
dependent but was significantly increased with attention (Figure 6C).

The excitatory firing rate was not altered by attention because of the
balance between two effects. The increased firing rate of CI cells
{\it increases} the amount of inhibition the excitatory cells receive
directly from CIs, but it also {\it decreases} the firing rate of SI
cells, which decreases the amount of SI inhibition to the excitatory
cells. The overall effect is not straightforward because the degree
of synchrony of the inhibitory inputs also varied, which by itself,
even in the absence of any rate changes, could modulate the
postsynaptic neuron's firing rate (see \citep{Tiesinga05}). The
precise balance depends on how much of the inhibitory input to the SI
cells comes from the CI cells. For the present parameter setting,
34\% comes from CI cells (Table \ref{tab2}). When the fraction of inputs
coming from the CI cells was increased, the firing rate of the
excitatory cells increased significantly with attention. This
parameter setting may thus be more appropriate for cortical areas
downstream of V1.

The nonlinearity present in the CRF for low contrast depends on the strength
of recurrent excitation. For low contrast, the firing rate elicited in
response to the stimulus grew {\it over time}, because of the recurrent
excitation, to reach its maximum 100 ms or more after response
onset. For this range of contrast values, the CRF increased steeply
with contrast. For higher contrast, or a higher unitary strength of
excitatory synapses, the firing rate reached its maximum value sooner
{\it in time}. The leveling-off of the rate of increase of the CRF was
associated with the firing rate reaching its maximum value shortly
after response onset and a change in the dynamics of the interneurons.
Specifically, in the attended state, the complex interneurons
increased their firing rate and oscillation frequency but their
precision was decreased (see below). In the non-attended state, the
simple inhibitory cells increased their firing rate.

Figure \ref{fig6}D-F shows the CRF for neurons that prefer a horizontal
orientation, which is orthogonal to the presented stimulus
orientation. The excitatory cells did not fire for any value of the
contrast (Figure \ref{fig6}D), whereas the simple inhibitory
cells fired but their firing rate decreased as a function of contrast
(Figure \ref{fig6}E). In the attended 
condition their CRF was shifted downward by approximately 4 Hz compared
with the non-attended condition. The complex cells labeled as
preferring horizontal behaved exactly the same as the ones shown in
Figure \ref{fig6}C because they were not orientation selective.

\begin{figure}
\centering
\epsfig{file=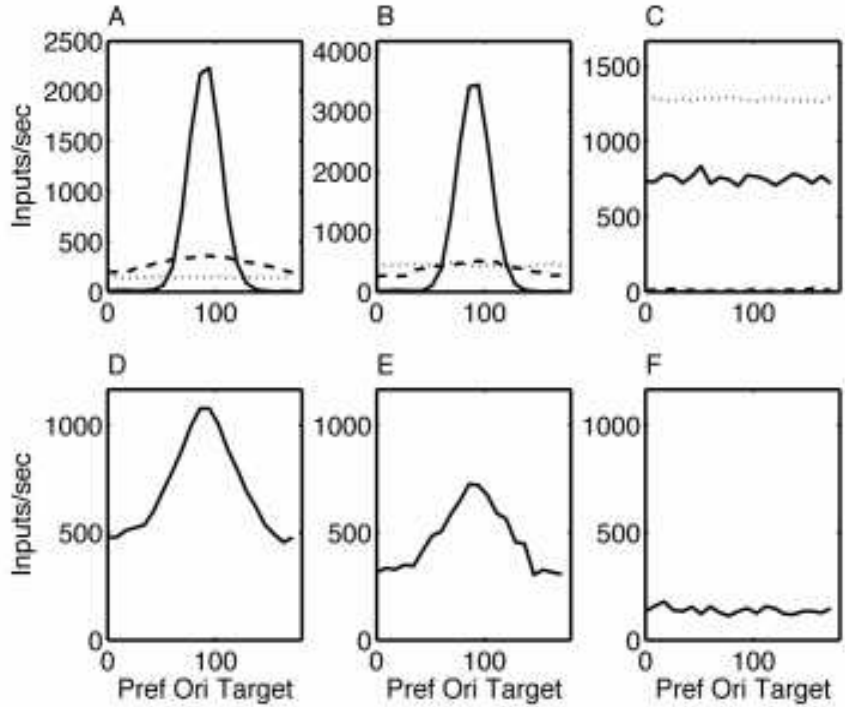, width=5in}
\caption{Tuning of the synaptic inputs to cortical neurons in response
  to the presentation of a vertical bar. We show the rate of inputs to
  (A) excitatory, (B) simple inhibitory and (C) complex inhibitory
  neurons from excitatory (solid lines), simple inhibitory (dashed
  lines) and complex inhibitory neurons (dotted lines) as a function
  of the neuron's preferred orientation. The responses were averaged
  across all neurons of a specific type in each column.  In the bottom
  three graphs the LGN inputs to (D) excitatory, (E) simple inhibitory
  and (F) complex inhibitory neurons are shown.
}
\label{fig7}
\end{figure}

We studied the tuning of synaptic drives from the LGN, SI, CI and E
populations by calculating for a given stimulus orientation the rate
of inputs to a neuron as a function of its preferred orientation
(Figure \ref{fig7}). Like the orientation tuning curves discussed before, these
curves can be reinterpreted as the input rate to one specific column
as a function of stimulus orientation. The excitatory neurons received
moderately tuned input from the LGN (Figure \ref{fig7}D), sharply tuned input
from other excitatory neurons (Figure \ref{fig7}A, solid line), and weakly
tuned input from simple inhibitory cells (Figure \ref{fig7}A, dashed line). The
input from complex inhibitory cells was not orientation-tuned (Figure
\ref{fig7}A, 
dotted line). The simple inhibitory cells received similarly tuned
inputs (Figure \ref{fig7}B and E), except that the rate of the LGN inputs was
lower because fewer LGN neurons project to the interneurons (Table \ref{tab2}),
whereas the rate of excitatory inputs was higher.  The inhibitory
inputs from complex cells dominated those coming from the simple
inhibitory cells. None of the synaptic inputs to complex cells were
orientation-tuned (Figure \ref{fig7}C and F).

\begin{figure}
\centering
\epsfig{file=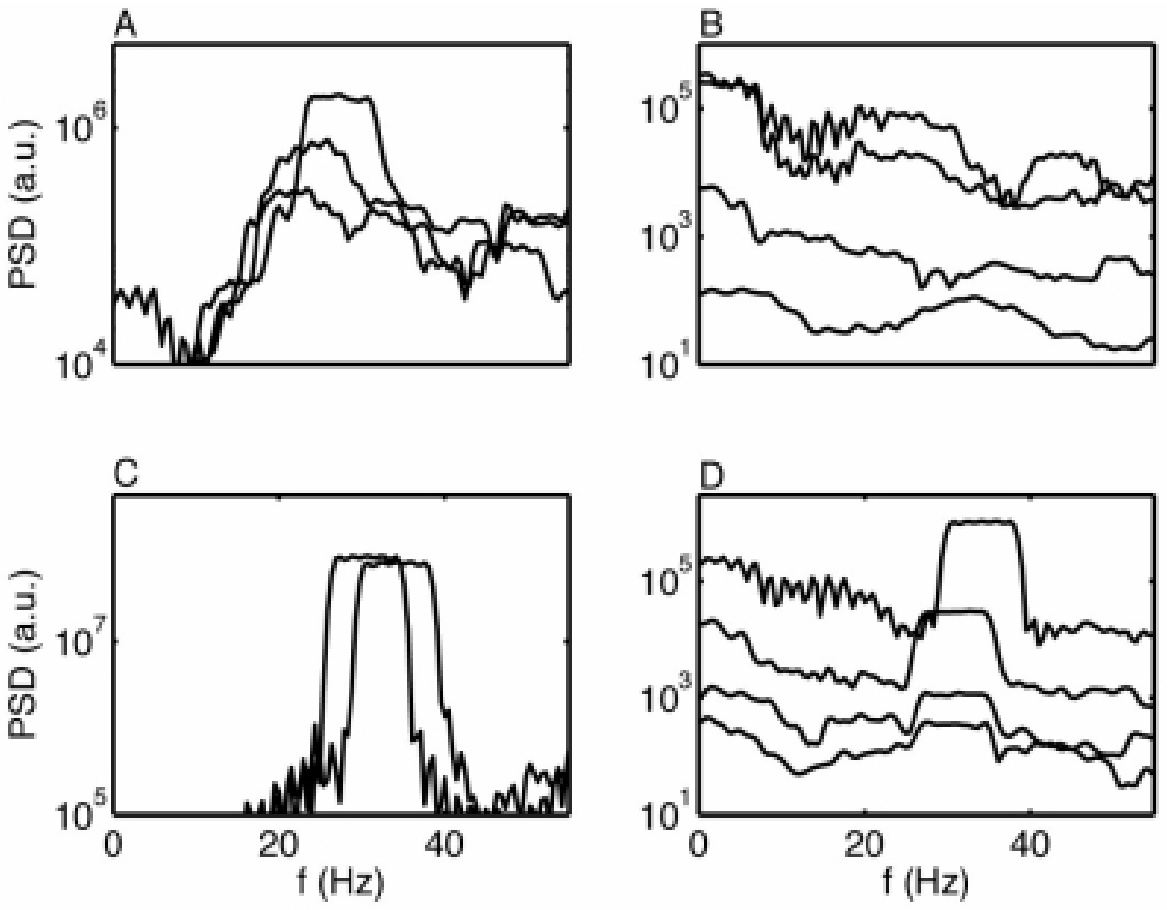, width=5in}
\caption{Gamma-frequency-range power in the estimated local field
  potential increased with contrast. We show the power spectrum
  density of the spike time histogram (eLFP) of the (A,C) complex
  inhibitory neurons and (B,D) excitatory neurons in the (A,B)
  non-attended state and (C,D) attended state. In each panel, power
  spectra are shown for different values of the contrast: (A), from
  bottom to top, 0, 35 and 100\%; (B), from bottom to top, 0, 25, 50
  and 100\%;  (C), left 0\% and right 100\%; (D) from bottom to top,
  0, 10, 25 and 100\%. The results are for a complex ring model in the
  attended state (as in Figure 5). The power spectrum was calculated
  across 600 samples of the spike time histogram with a 1 ms time
  resolution, and averaged across 5 tapers with a bandwidth of NW=3.
}
\label{fig8}
\end{figure}

We estimated the temporal modulation of synchrony using a multi-taper
spectrogram of the spike time histogram of the E, SI and CI
populations. The LFP is hypothesized to reflect the synaptic currents
to neurons in the neighborhood of the electrode \citep{Liu06}. Hence,
when the spike time histograms for each population are 
filtered by the corresponding synaptic time constants and added
together with the appropriate weights, they might behave similar to
the LFP. We therefore refer to the (unfiltered) spike time histogram
as the estimated LFP, eLFP for short. In the non-attended condition
(Figure \ref{fig8}A), complex cells show power in the frequency range between
22 to 28 Hz which grows in amplitude and frequency as a function of
contrast. The corresponding oscillations are only visible in the spike
time histogram for the 100\% contrast stimulus (Figure \ref{fig4}C). Power at
this frequency was absent or barely detectable in the simple
inhibitory and excitatory cells (Figure \ref{fig8}B).  In the attended state
(Figure \ref{fig8}C and D), the complex cells were synchronized during the
entire period, including when there was no stimulus present. During
stimulus presentation the synchrony was maintained, but for higher contrast
the oscillation frequency increased whereas its power decreased
slightly (Figure \ref{fig8}C). The simple inhibitory and excitatory cells
followed the CI generated rhythm when they spiked, that is, during
the stimulus period. The gamma-frequency range power in the eLFP thus
increased with contrast (Figure \ref{fig8}D).

\begin{figure}
\centering
\epsfig{file=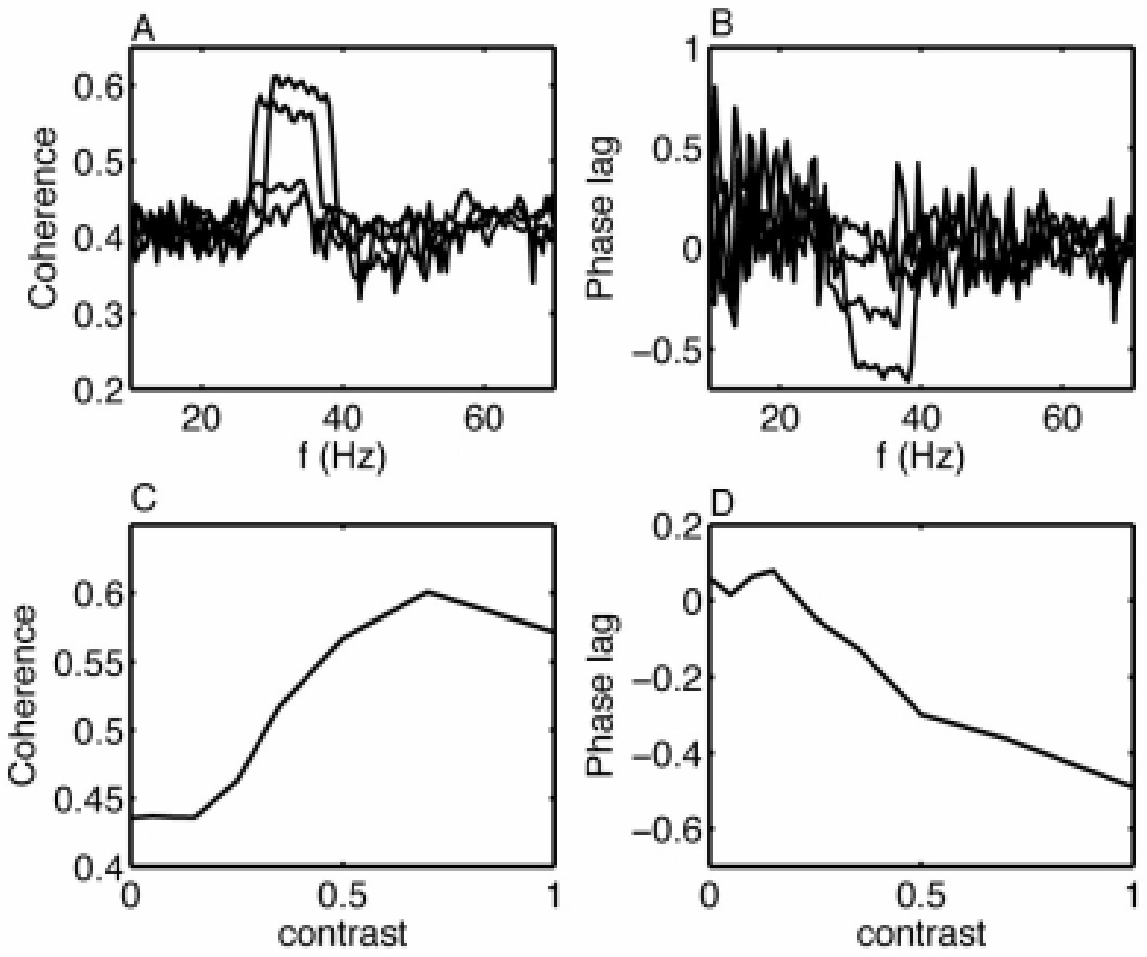, width=5in}
\caption{The gamma-frequency-range coherence between the estimated
  local field potential and the excitatory neurons increased with
  contrast. The (A) coherence and (B) relative phase between the spike
  time histogram of inhibitory neurons and the spike trains of
  excitatory neurons as a function of frequency. A negative phase
  means that the excitatory neurons are ahead of the inhibitory
  neurons. The (C) coherence and (D) relative phase in the 28 to 36 Hz
  frequency band is plotted as a function of stimulus contrast. The
  results are for a complex ring model in the attended state (as in
  Figure \ref{fig5}) and are averaged across a hundred spike trains of
  excitatory neurons. The coherence was calculated across 600 samples
  of the spike time histogram with a 1 ms time resolution, and
  averaged across 5 tapers with a bandwidth of NW=3. 
}
\label{fig9}
\end{figure}

The coherence between the excitatory and inhibitory population was
quantified using the multi-taper coherency \citep{Mitra99,Jarvis01}. We
calculated the coherencies between the 
spike time histograms (not shown) as well as between the excitatory
spike trains and the eLFP of the CI cells (Figure \ref{fig9}).  The latter
measure is similar to the spike field coherence used in
\citep{Fries01}. The coherence is a complex quantity -- its absolute value 
represents the strength of the coherence (Figure \ref{fig9}A and C) and its
phase is proportional to the delay between the activity of the two
populations (Figure \ref{fig9}B and D). During the stimulus period in the
attended state, the coherence showed a clear peak for frequencies
between 28 and 40 Hz (Figure \ref{fig9}A) with the CIs lagging the excitatory
neurons, as is indicated by the negative phase (Figure \ref{fig9}B). The
coherence grew with contrast, leveling off at 70\% contrast (Figure
\ref{fig9}C). Likewise, the phase lag increased, becoming more negative with
contrast (Figure \ref{fig9}D).  In the non-attended state, the level of
coherence was not significant when tested at a p-value of
0.05. However, in the 22 to 32 Hz frequency range, there was coherence
between the eLFPs of the excitatory and CI neurons (not shown).

\begin{figure}
\centering
\epsfig{file=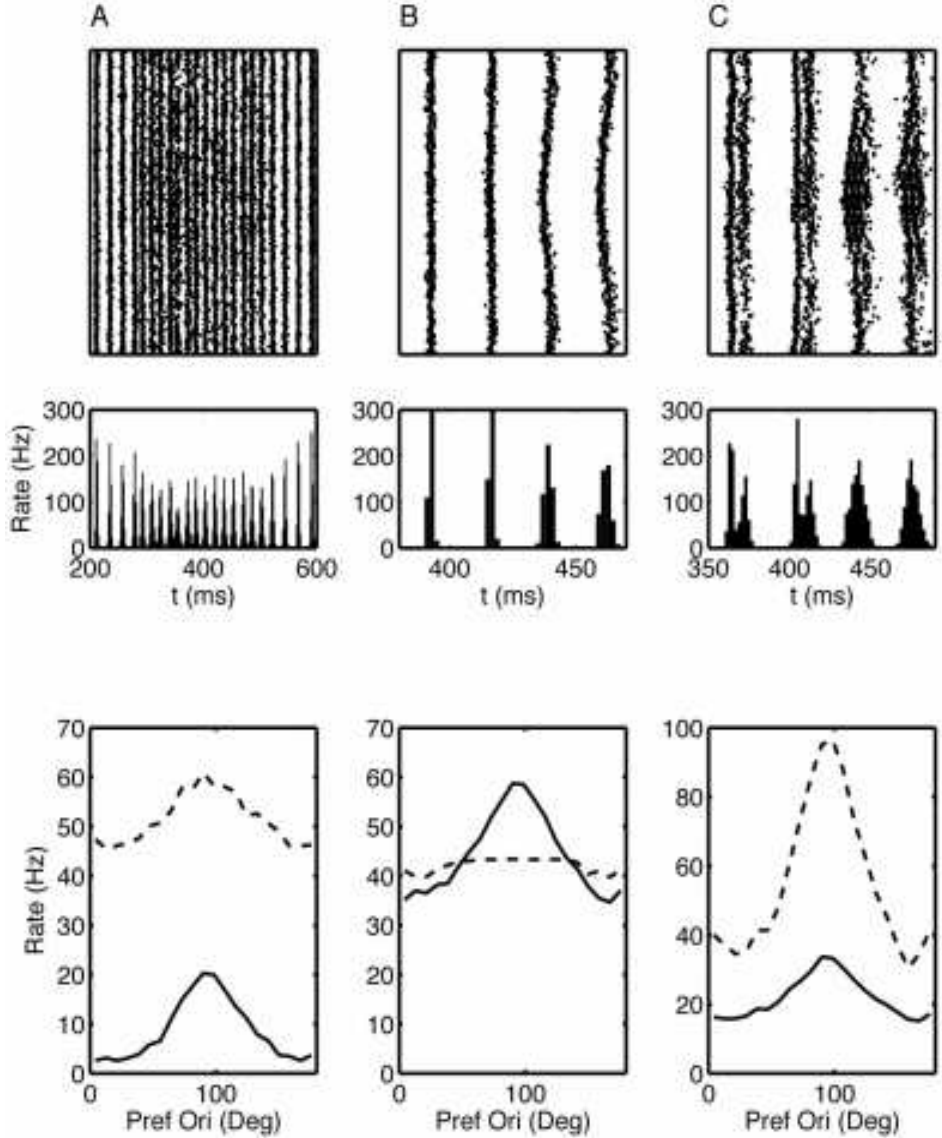, width=5in}
\caption{Alternative mechanisms for obtaining synchrony modulations in
  the standard ring model. In each panel, we show (top) the
  rastergram, (middle) the histogram of simple inhibitory cells and
  (bottom) the firing rate in response to a vertical bar as a function
  of the preferred orientation for excitatory (solid lines) and simple
  inhibitory cells (dashed lines). We present simulations for three
  cases, when stimulus presentation increased the oscillation
  frequency (A), reduced the latency (B), or increased the number of
  spikes per cycle (C) of neurons with a preferred orientation that
  matched the stimulus orientation. The parameter values are listed in
  Table \ref{tab3}.
}
\label{fig10}
\end{figure}

Is it possible to obtain synchrony modulation without adding complex
cells to the network? Theoretical and computational studies have
identified two different ways of obtaining synchronous oscillations in
the gamma frequency range \citep{Wang96,White98,Brunel00,Tiesinga01,
Aradi02,Bartos02,Borgers03,Brunel03,Hansel03,Borgers05,Vida06}. In the
first one, "PING" \citep{Borgers03}, a synchronous volley of spikes from
excitatory cells elicits a volley of spikes from the inhibitory cells,
which shuts down the network for approximately a gamma period, 25 ms,
after which 
the cycle starts anew. For this rhythm to be stable, the inhibitory
cells should not be able to spike before the synchronous excitatory
volley arrives \citep{Borgers05}. In the network studied
here, excitatory and inhibitory cells are both driven by almost
simultaneous LGN inputs that can make them spike even without
intracortical excitation, suggesting that the PING mechanism might not
be effective in this network architecture. In the second one, "ING",
the inhibitory network synchronizes by way of mutual inhibition
\citep{Wang96}. This rhythm requires about one hundred neurons and a 
high degree of interconnectedness \citep{Wang96,Golomb00}. Furthermore,
it can only exist when the random 
background activity (noise) and the degree of heterogeneity
(differences in intrinsic excitability) are small enough \citep{Wang96,
  White98, Tiesinga00}. It is the 
heterogeneity that posed the main problem in our explorations of the
standard ring model. In Figure \ref{fig7}B and D the rates of intracortical and
LGN inputs to interneurons are shown. The interneurons whose preferred
orientation matched the stimulus orientation received much stronger
inputs from the excitatory neurons and LGN neurons than did those whose
preferred orientation was different from the stimulus orientation. The
intracortical connections contributed more to the heterogeneity than
the LGN inputs. We explored how far we could get, using only simple
cells, towards our goal of finding a network that is synchronous in
the prestimulus period and remains synchronous after the onset of the
stimulus. The number of inhibitory cells was tripled to 63 per column
and the number of inhibitory synapses onto each interneuron was
increased to 128. Since the intracortical inputs caused the most
heterogeneity, the intracortical connections were either made
independent of the difference in orientation preference between the
presynaptic cell and the postsynaptic cell or their unitary strength
was reduced. In 
addition, the unitary strength of the thalamocortical synapses was
reduced slightly. We found three ways in which the increase in
heterogeneity associated with stimulus onset could be absorbed by the
network without losing synchrony. First, the network could increase
its oscillation frequency (Figure \ref{fig10}A). Second, it could decrease the
latency of the interneurons whose preferred orientation matched that
of the stimulus without changing the firing rate (Figure \ref{fig10}B). Third,
it could increase the number of spikes the interneuron produces on
each cycle without changing the oscillation (Figure \ref{fig10}C). The
parameter settings are summarized in Table \ref{tab3}.

\begin{table}
\centering
\begin{tabular}{|l|c|c|c|c|c|c|}
\hline
  & E to E & E to I & I to E & I to I & $I_E$ & $I_I$ \\
  &($\mu\textrm{S/cm}^2$)& ($\mu\textrm{S/cm}^2$) &
  ($\mu\textrm{S/cm}^2$) & ($\mu\textrm{S/cm}^2$) & 
  ($\mu\textrm{A/cm}^2$) & ($\mu\textrm{A/cm}^2$) \\
\hline
{\bf A} & 2 & 20 & 5 & 3 & 0 & 2 \\
\hline
{\bf B} & 1.2 & 1 & 2 & 3 & 1.1 & 2 \\
\hline
{\bf C} & 2 & 20 & 5 & 3 & 0.7 & 0.5 \\
\hline
\end{tabular}
\caption{Parameters for the simulations shown in Figure \ref{fig10}. The
  parameters are as in the standard ring model, except that there were
  63 inhibitory cells (increased from 21) and each inhibitory cell received
  128 inhibitory synapses. The thalamocortical synapses onto
  excitatory cells had a unitary strength of 12$\mu\textrm{S/cm}^2$, those on
  inhibitory cells were 7$\mu\textrm{S/cm}^2$. $I_E$ and $I_I$ are the
  currents to the excitatory and inhibitory neurons, respectively.}
\label{tab3}
\end{table}

 In case 1 (Figure \ref{fig10}A), the interneurons received a large
 depolarizing drive and the mutual inhibition was strong. As a result,
 the interneurons could synchronize without strong LGN inputs. The
 intracortical connections were made untuned, that is, the connection
 probability did not depend on the difference in preferred orientation
 between the presynaptic neuron and the postsynaptic neuron. When the
 vertical bar came 
 on, it increased the rate of the interneurons with a preferred
 orientation close to $90^{\circ}$. These neurons led the oscillation,
 thus increasing its frequency. Because of the strong mutual
 inhibition the interneurons with preferred orientations farther from
 $90^{\circ}$ remained entrained to the sped-up rhythm, but they occasionally
 skipped a cycle. The modulation of interneuron firing
 rate with preferred orientation was weak: the dynamic range -- the
 difference between the highest and lowest firing rate --  was about
 15 Hz. This modulation was on top of a high baseline rate of about 46
 Hz. The dynamic range of the excitatory neurons was similar, around
 18 Hz, with a baseline rate of 2.5 Hz.

For case 2 (Figure \ref{fig10}B), the interneurons still received a large
depolarizing current, but now the excitatory neurons also received a
depolarizing current. The intracortical connections were made
orientation-selective again, but the strength of all intracortical
connections was reduced except for the strength of the mutual
inhibition, which was kept the same. For these parameters, the
excitatory neurons fired at a high rate even before stimulus onset and
the network was synchronized. When the vertical bar came on, it
increased the firing rate of the excitatory neurons with an
orientation preference close to $90^{\circ}$. However, because the
strength of the excitatory synapses on the interneurons was weak, the
stimulus onset did not increase interneuron firing rate, rather it
decreased the spiking latency of interneurons preferring vertical bars
compared with those that preferred different orientations. The
inhibitory cells were not orientation-selective in the classical
sense since their firing rate did not change with stimulus
orientation. The dynamic range of the excitatory neuron was similar to
that in case 1, about 24 Hz, but it sat on top of a much higher
baseline rate of 35 Hz.

For case 3 (Figure \ref{fig10}C), the depolarizing current to excitatory and
inhibitory neurons was reduced. The unitary strength of the
intracortical connections was returned to their values for case 1, but
they remained orientation-selective as in case 2. Before stimulus
onset, the excitatory and inhibitory neurons were synchronized with an
oscillation frequency of about 25 Hz. The excitatory neurons fired one
spike on each cycle and led the interneurons, which fired two spikes
on each cycle. When the vertical bar stimulus came on, the
interneurons with an orientation preference close to $90^{\circ}$
increased their firing rate and spiked three times on each cycle,
whereas the interneurons preferring orthogonal orientations reduced
their rate to about one spike per cycle. The oscillation frequency
increased slightly to 28.6 Hz. The excitatory neurons preferring
vertical stimuli also increased their firing rate, but they still
spiked at most once on each oscillation cycle. The dynamic range of
the interneurons was large, 66 Hz, on top of a baseline rate of about
31 Hz, whereas the excitatory neurons had a small dynamic range of
about 19 Hz on top of a 15 Hz baseline. 

The outcome of this exploration is that it is possible to
modulate synchrony. However it is not possible, at the same time, to
significantly sharpen the orientation tuning of cortical neurons
compared with that of the LGN inputs by way of recurrent excitation
and to have a large dynamic range of the firing rate.

\section{Discussion}

Multiple features of a visual stimulus are represented in the neural
activity which it elicits in the visual pathway. These features include
orientation, luminance contrast and whether the stimulus is in the focus of
spatial attention. The most frequently used measure of neural activity
is firing 
rate. Consider an orientation-selective neuron that has a baseline
rate of 5 Hz, responds to a stimulus that matches its
preferred orientation with 30 Hz if the stimulus is at 50\% contrast
and with 60 
Hz if it is at 100\% contrast. If a rate of 25 Hz is measured, what
could the stimulus have been? It could have been a high contrast
stimulus with an orientation different from the neuron's preferred
orientation, or a low contrast stimulus with the neuron's preferred
orientation, or a low contrast stimulus that was in the focus of
attention. Or it could be that no stimulus was presented at all, but
that the neuron's baseline firing rate increased because of general
arousal. How does the visual system disambiguate the different
possible meanings of such a firing rate response? And, are the changes
in activity due to stimulus identity, stimulus strength and the focus of
attention generated by distinct circuitry? The answer to the first
question is that the population activity must be used,
with the identity of the neurons that are most strongly
activated and the degree of coherence/correlation between them being
the most significant aspect of the population activity. So
far, electrophysiological recordings in non-human primates have only
offered a glimpse of the richness of the population
response. Nevertheless it has become clear that in the primary visual
cortex of anesthetized macaque monkeys the LFP power in the gamma
frequency range increases with contrast \citep{Henrie05},
whereas the gamma-frequency-band coherence between a neuron's spike
train and the LFP increases with attention
\citep{Fries01,Bichot05}. Thus, stimulus strength and the focus of
attention modulate neural correlations \citep{Salinas01}. There is
no definite answer to the second question, primarily because there is
a lack of models and experiments that have addressed the issue of how
attention, contrast and stimulus identity interact. This study is our
first attempt to resolve this issue. 

Our goal was to determine how the effects of attention could be
incorporated in a model that produces contrast-invariant
orientation-selective responses. The model needed to reproduce the
following observations. First, contrast-invariant orientation tuning
curves should have a half width at half height between $10^{\circ}$
and $30^{\circ}$ \citep{Gur05}. Second, the power of the local field
potential (LFP) in the gamma-frequency should increase with contrast
\citep{Henrie05}. Third, the width of the orientation curve
should not change with attentional condition \citep{McAdams99}. Fourth,
attention should induce only small changes in the 
firing rate of neurons in responding to a simple stimulus
\citep{McAdams05}. Fifth, attention should modulate the power of the
LFP, as well as the coherence of spike trains with the LFP, in the gamma
frequency range \citep{Fries01, Bichot05, Taylor05}.  We show that it
is easiest to assume the existence of 
interneurons with complex receptive field characteristics, that are
weakly or not at all orientation-selective and that show a modest increase in
firing rate with contrast, but that are strongly modulated by
attentional state. We have not conclusively shown that similar
dynamics could not be achieved using only excitatory and inhibitory
cells with simple receptive field characteristics, but so far we have
not been successful in finding such a network that behaves
appropriately. The reason for this is that the inhibitory neurons had to serve
two functions. They needed to sharpen orientation-selectivity and they
needed to synchronize. Synchrony develops in interneuron networks when
(1) there are enough interneurons, about one hundred (Wang and Buzsaki,
1996), (2) they are well connected, about sixty synapses per
neuron \citep{Wang96}, (3) the level of background activity,
noise, is low enough \citep{Tiesinga00} and (4) the degree of
heterogeneity is low enough \citep{Wang96, White98}. Not only does the
variability in intrinsic properties of neurons 
contribute to the heterogeneity, but the inhomogeneous activation of
neurons also contributes because of their different preferred orientations. Our
solution was to assign a different interneuron to each function.  In
experiments, two types of interneurons were found in layer 4 of cat
primary visual cortex in approximately equal proportions
\citep{Hirsch03}. One had a simple receptive field and was 
orientation-selective, whereas the other had a complex receptive field
and was not orientation-selective. In a recent model these complex
interneurons were shown to be useful for contrast-invariant
orientation tuning \citep{Laur03}. The question remains
whether these complex interneurons also exist in layer 4 of primate
visual cortex. The mechanism proposed here predicts that there is a
pool of interneurons with complex RF characteristics that increases
their rate with attention and another pool of interneurons with simple
RFs that decreases their rate with attention. 

We studied the behavior of the model in two operating regimes. In the
first, the activity before stimulus onset was asynchronous, but the
onset of a sufficiently high contrast stimulus led to
weak synchronous oscillations in the spike time histogram of complex
inhibitory cells. Hence, the gamma-frequency power in the eLFP associated
with the CI cells increased with stimulus contrast, consistent with
the previously mentioned experimental results \citep{Henrie05}. We
propose that this parameter setting represents the 
non-attended state.  In the second, the complex cell activity was
synchronous before and also during stimulus presentation, but the
oscillation frequency increased during the stimulus presentation when
the contrast was high enough. The excitatory cells were also
synchronized. Their rate was low for low contrast, but it increased
with contrast. As a result, the coherence between excitatory activity
and the eLFP (from CI cells) increased with contrast. We propose that
this parameter setting represents the attended state. The network can
be switched from the non-attended to attended state solely by
increasing the drive to the complex interneurons. This drive could be
supplied by various subcortical \citep{Freund92} and
cortical feedback projections \citep{Gonchar99, Gonchar03}. 

The preceding discussion has focused on the modulation of synchrony by
attention. In cortical areas V2, V4 and MT, changes of firing rate
with spatial attention have been reported \citep{Luck97, McAdams99,
Treue99, Reynolds00, Willi06}.  In V1, no clear evidence for firing
rate changes with attention have been found when using simple stimuli
\citep{McAdams05} (but see \citep{Roelf04, Kha06}). However,
attention may cause changes in synchrony in V1, which could lead to
firing rate increases downstream \citep{Tiesinga04}. Hence, our
parameter settings for the V1 model were chosen such that
there was little change in the CRF of excitatory neurons whose
preferred orientation matched the stimulus orientation. This was
achieved by tuning the number of synapses from CI to SI cells such
that the increase in inhibition from CI cells to the excitatory cells
with attention was balanced by the decrease in inhibition from SI
cells.  An increase in excitatory activity with attention, as is
observed in V4, can be obtained in this type of network by decreasing
the fraction of synapses on SI neurons coming from CI cells. 

We used the ring model to generate orientation selective neurons. The
ring model has experimental support \citep{Tsodyks99, Kenet03} and has
been used previously in large-scale model 
simulations \citep{Somers95}. There are alternative models, some
of which require strong recurrent excitation and inhibition
\citep{McLaughlin00, Marino05}, others of which are based primarily on
the selectivity of thalamic inputs \citep{Troyer98, Ferster00}. The
actual situation may lie 
between these two extreme cases \citep{Monier03, Teich06}. We have
chosen the ring model as the first case to study, but 
it is of interest to determine how the considerations presented here
apply to the alternative models. 

Experiments with multi-electrode arrays in awake behaving primates are
presently being conductedin a number of labs, making it possible to
record from multiple 
neurons simultaneously during an attention-demanding task. The model
can potentially be useful for these experiments since it predicts how
the coherence between neurons changes with attention and contrast and
how it depends on the difference in stimulus preference between the
neurons.

{\bf Acknowledgements:} This work was supported in part by start-up
funds from the University of North Carolina at Chapel Hill and by the
Human Frontier Science Program.

\bibliographystyle{elsart-harv}

\bibliography{soltesz.bib}

\end{document}